\documentclass[aps,prb,preprint,showpacs]{revtex4}

\usepackage{amsmath}
\usepackage{bm}
\usepackage{graphicx}

\begin{document}

\title{Electronic transport through a
parallel-coupled triple quantum dot molecule: Fano resonances and
bound states in the continuum}

\author{M. L. Ladr\'on de Guevara}
\author{P. A. Orellana}

\affiliation{Departamento de F\'{\i }sica, Universidad Cat\'{o}lica
del Norte, Casilla 1280, Antofagasta, Chile}

\begin{abstract}
In this article we study electronic transport through a triple
quantum dot molecule attached in parallel to leads in presence of a
magnetic flux. We have obtained analytical expressions of the linear
conductance and density of states for the molecule in equilibrium at
zero temperature. As a consequence of quantum interference, the
conductance exhibits one Breit-Wigner and two Fano resonances, which
positions and widths are controlled by the magnetic field. Every two
flux quanta, there is an inversion of roles of the bonding and
antibonding states. For particular values of the magnetic flux and
dot-lead couplings, one or even both Fano resonances collapse and
bound states in the continuum (BIC's) are formed. We examine the
line broadenings of the molecular states as a function of the
Aharonov-Bohm phase around the condition for the formation of BIC's,
finding resonances which keep extremely narrow against variations of
the magnetic field. Moreover, we analyze a molecule of $N$ quantum
dots in the absence of magnetic field, showing that certain
symmetries lead to a determinate quantity of simultaneous BICs.
\end{abstract}

\date{\today}


\maketitle

\section{Introduction}

Electron transport through quantum dot configurations has been a
subject of permanent interest in the last years. Since quantum dots
electrons are confined in all three spatial dimensions, they are
also called ``artificial atoms'',\cite{atoms} and two or more
quantum dots can be coupled to form ``artificial molecules''.
Tunneling through a diatomic artificial molecule in a configuration
in series has been extensively studied, both theoretically and
experimentally.\cite{blick,ziegler,vanderwiel,golovach} In Refs.
\onlinecite{waugh1,waugh2} Waugh {\em et al.} reported the
observation of peak splitting in the conductance through double and
triple quantum dot molecules. The formation of  band structures in
finite one-dimensional arrays of quantum dots has been also
discussed.\cite{crystal,zeng}

A distinctive feature of electron tunneling through quantum dots
is the retention of the quantum phase coherence. For this reason,
multiple connected geometries involving quantum dots exhibit
quantum interference phenomena, such as the Fano
effect,\cite{kang,gores,kobayashi} which arises from the
interference between a discrete state and the
continuum.\cite{fano} Several works have been concerned with the
study of transmission through a parallel-coupled double quantum
dot molecule embedded in an Aharonov-Bohm
interferometer.\cite{kang2,bai,orellana,aldea,lu} This is
characterized by the formation of a tunable Fano resonance in the
conductance spectrum. This resonance is associated to a long-lived
molecular state, the position and lifetime of which are controlled
by the magnetic field. For some particular values of the magnetic
flux, that molecular state decouples completely from the
leads,\cite{orellana} becoming a ``bound state in the continuum"
(BIC). Such a resonant state with infinite lifetime was called
``ghost Fano resonance'' in Ref. \onlinecite{lldg}.

The existence of bound states embedded in the continuum was early
proposed by von Neumann and Wigner for certain spatially oscillating
attractive potentials for a one-particle Schr\"{o}dinger
equation.\cite{boundstate1} Much later, Stillinger and Herrick
generalized von Neumann's work and analyzed a two-electron problem,
where BICs were formed despite the interaction between
electrons.\cite{stillinger} The occurrence of BICs was discussed in
a system of coupled Coulombic channels and, in particular, in an
Hydrogen atom in a uniform magnetic field\cite{friedrich}. The
authors interpreted the formation of these states as result of
interference between resonances of different channels.

Bound states in the continuum have also shown to be present in
electronic transport in mesoscopic structures. There are theoretical
works showing the formation of these states in a four-terminal
junction\cite{schult} and in a ballistic channel with intersections.
\cite{zhen-li} Experimental evidence of BICs was reported by Capasso
{\em et al.}\cite{capasso} in semiconductor heterostructures grown
by molecular beam epitaxy. Bound states in the continuum have been
discussed little in the context of quantum dots. In Ref.
\onlinecite{nockel} it was studied the ballistic transport through a
quantum dot and demonstrated the possibility of a classical
analogous of BICs. These states have also been found in a curved
waveguide with an embedded quantum dot\cite{olendski} and they arise
in transport through a double quantum dot in series with two
relevant levels in each dot.\cite{rotter}

In this article we study electron transport through a parallel
triple quantum dot molecule embedded in an Aharonov-Bohm
interferometer connected symmetrically to leads, and we focus in
the formation of bound states in the continuum. It is assumed the
system is in equilibrium at zero temperature, and
electron-electron interactions are neglected. We find that by
choosing appropriately the dot-lead tunneling couplings, up to two
of the three molecular states may simultaneously decouple of the
leads, becoming BICs. We analyze the role played by the magnetic
flux in the participation of the molecular states in transmission,
and in particular in the survival of the bound states in the
continuum. We observe that different regimes of transmission can
be reached by varying the magnetic field. With a period of two
flux quanta, the roles of the antibonding and bonding states are
interchanged in the conductance spectrum. On the other hand,
either one or two BICs are periodically formed as the magnetic
flux is varied, and the broadenings of these states result
extremely robust to variations of the flux over a wide range. This
robustness is not present in the double molecule.\cite{kang}
Finally, we make a brief analysis of an array of $N$ quantum dots,
with $N$ arbitrary, showing that  certain symmetries guarantee the
formation of a determined number of BICs. We give detailed
examples of the cases with $N=4$ and $5$.

The paper is organized as follows. In Sec.\ref{model} we introduce
the Hamiltonian of the system, and we develop the equation of
motion approach for the Green's functions, in order to obtain
expressions for the total density of states and linear conductance
at zero temperature. We also examine the conditions for the
formation of BICs. In Sec.\ref{results} we present the results for
the conductance and density of states for two different set of
parameters. We also analyze the line broadenings of the molecular
states as a function of the Aharonov-Bohm phase. We discuss the
$N$ quantum dot molecule in Sec.\ref{Ndots}, and in
Sec.\ref{conclusions} we give our concluding remarks.

\section{Model}\label{model}

\begin{figure}
\begin{center}
        \includegraphics{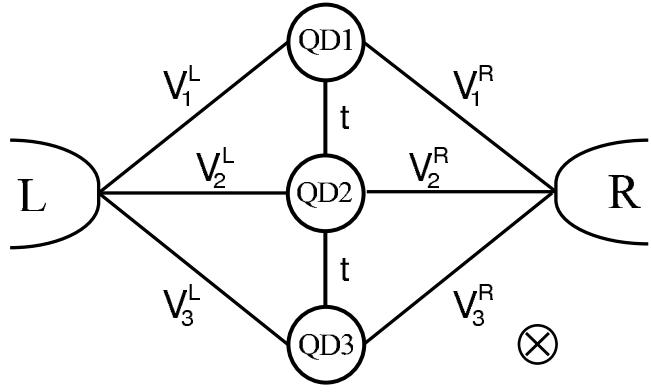}
\end{center}
\caption{Triple quantum dot molecule coupled in parallel to leads.}
\label{fig1}
\end{figure}

We consider three single-level quantum dots forming a triple quantum
dot molecule coupled in parallel to two leads, as shown in Fig.
\ref{fig1}. The system is modeled by a noninteracting Anderson
Hamiltonian, which can be written as
\begin{equation} H=H_{m}+H_{l}+H_{I},  \label{eq-1}
\end{equation}
where $H_{m}$ describes the dynamics of the isolate molecule,
\begin{equation}
H_{m}=\sum_{i=1}^{3}\omega _{i}d_{i}^{\dag }d_{i}-t(d_{1}^{\dag
}d_{2}+d_{2}^{\dag }d_{1})-t(d_{2}^{\dag }d_{3}+d_{3}^{\dag }d_{2}),
\label{eq-2}
\end{equation}
\noindent where $\omega _{i}$ is the level energy of dot $i$, $d_{i}$ $%
(d_{i}^{\dagger })$ annihilates (creates) an electron in dot $i$,
and $t$ is the interdot tunneling coupling. $H_{l}$ is the
Hamiltonian for the noninteracting electrons in the left and right
leads
\begin{equation}
H_{l}=\sum_{{k}\in L,R} \omega _{k} c_{k}^{\dag}c_{k}, \label{eq-3}
\end{equation}
where $c_{k}$ $(c_{k}^{\dag })$ is the annihilation (creation)
operator of an electron of quantum number $k$ and energy $\omega
_{k}$ in the contact $L$ or $R$. The term $H_{I}$ accounts for the
tunneling between dots and leads,
\begin{eqnarray}
H_{I} &=&\sum_{i=1}^{3}\sum_{k \in L} V^L_{i}d_{i}^{\dag}
c_{k}+\mbox{h. c.} + \sum_{i=1}^{3} \sum_{{k} \in R}
V^R_{i}d_{i}^{\dag}c_{k}+\mbox{h. c.} \nonumber \\
\label{eq-4}
\end{eqnarray}
with $V^{L(R)}_{i}$ the tunneling matrix element connecting the
$i-th$ dot with the left (right) lead, assumed independent of $k$.
For simplicity, we assume that the magnitudes of these matrix
elements are such as $|V^L_{1}|=|V^R_{1}|\equiv V_1$,
$|V^L_{2}|=|V^R_{2}|\equiv V_2$, and $|V^L_{3}|=|V^R_{3}|\equiv
V_3$. In presence of a magnetic field, and in the symmetric gauge,
the tunnel matrix elements can be written in the form
\begin{eqnarray}
V^{L}_{1}&=&V_{1}e^{-i\phi/4}, \quad V^{R}_{1}=V_{1}e^{i\phi/4},
\nonumber\\
V^L_{2}&=&V^R_{2}=V_2, \nonumber \\
V^{L}_{3}&=&V_{3}e^{i\phi/4}, \quad V^{R}_{3}=V_{3}e^{-i\phi/4},
\label{eq-5}
\end{eqnarray}
with $\phi=2\pi\Phi/\Phi_0$, the Aharonov-Bohm phase, where
$\Phi_0=h/e$ is the flux quantum.

The linear conductance at zero temperature is given by the Landauer
formula
\begin{equation}
G=\frac{2e^{2}}{h}T(\varepsilon_F), \label{eq-6}
\end{equation}
where $T(\omega)$ is the total transmission. To obtain $G$
explicitly we use the equation of motion approach for the Green's
functions.\cite{greens} The transmission can be expressed in terms
of the retarded and advanced Green's functions
$\textbf{G}^{r/a}(\omega)$ as
\begin{equation}
T(\omega)=tr\{\mathbf{G}^{a}(\omega )\mathbf{\Gamma }^{R}\mathbf{G%
}^{r}(\omega )\mathbf{\Gamma }^{L}\},  \label{eq-7}
\end{equation}
where $\mathbf{G}^r(\omega)$ is defined by
\begin{equation}
G_{ij}^{r}(t)=-i\theta (t)\langle \{d_{i}(t),d_{j}^{\dagger
}(0)\}\rangle ,\quad i,j=1,2,3,  \label{eq-10}
\end{equation}
with $\theta (t)$ the step function. $\mathbf{G}^a(\omega)$ is given
by $G_{ij}^{a}=[G_{ji}^{r}]^*$, and ${\mathbf {\Gamma}}^{L,R}$ are
matrices describing the coupling between the quantum dots and the
left and right leads, the matrix elements of which are
\begin{equation}
\Gamma^{L(R)}_{ij}=2\pi\sum_k V^{L(R)}_{ik}[V^{L(R)}_{jk}]^{*}
\delta(\omega-\omega_k), \quad \mbox{$i,j=1,2,3$}. \label{eq-8}
\end{equation}
With the use of Eq. (\ref{eq-5}), ${\mathbf {\Gamma}}^{L,R}$ can be
written as
\begin{equation}
{\mathbf \Gamma}^{L,R}=\left(
\begin{array}{lll}
\;\;\;\; \gamma_{11} &  e^{\mp i \phi/4}\gamma_{12} & e^{\mp i \phi/2}\gamma_{13} \\
e^{\pm i \phi/4}\gamma_{21} & \;\;\;\; \gamma_{22} & e^{\mp i \phi/4}\gamma_{23}\\
e^{\pm i \phi/2}\gamma_{31} &  e^{\pm i \phi/4}\gamma_{32} &
\;\;\;\; \gamma_{33}  \label{eq-9}
\end{array}
\right),
\end{equation}
with $\gamma_{ij}\equiv\Gamma^{L}_{ij}=\Gamma^{R}_{ij}$, where
$\Gamma^{L,R}_{ij}$ are obtained from (\ref{eq-8}) for zero magnetic
flux.

The electronic properties of the configuration can be studied from
the total density of states. This quantity is given by
\begin{equation}
\rho(\omega)=\sum_{i=1}^3\rho _{i}(\omega )=-\frac{1}{\pi
}\sum_{i=1}^3{\mbox Im}G_{ii}^{r}(\omega ),\quad i=1,2,3,
\label{eq-11}
\end{equation}
where $G^r(\omega)$ is the retarded Green's function.

Hereafter we assume $\omega_1=\omega_2=\omega_3\equiv \omega_0$. We
make the following transformation of the quantum-dot operators
\begin{eqnarray}
\bar{d}_1&=&\frac{1}{2}(d_1+\sqrt{2}d_2+d_3), \nonumber \\
\bar{d}_2&=&\frac{1}{\sqrt{2}}(d_1-d_3), \nonumber \\
\bar{d}_3&=&\frac{1}{2}(d_1-\sqrt{2}d_2+d_3), \label{eq-12}
\end{eqnarray}
so that the Hamiltonian of the isolated molecule becomes diagonal
\begin{equation}
\bar{H}_m=(\omega_0+\sqrt{2}t)\bar{d}^\dag_1\bar{d}_1+\omega_0
\bar{d}_2^\dag\bar{d}_2
+(\omega_0-\sqrt{2}t)\bar{d}^\dag_3\bar{d}_3, \label{eq-13}
\end{equation}
and the Hamiltonian describing the coupling between the molecule and
the leads takes the form
\begin{eqnarray}
\bar{H}_{I} &=&\sum_{i=1}^{3}\sum_{k \in L}
\bar{V}^L_{i}\bar{d}_{i}^{\dag} c_{k}+\mbox{h. c.} + \sum_{i=1}^{3}
\sum_{{k} \in R}
\bar{V}^R_{i}\bar{d}_{i}^{\dag}c_{k}+\mbox{h. c.}, \nonumber \\
\label{eq-14}
\end{eqnarray}
where
\begin{eqnarray}
\bar{V}^{L,R}_{1}&=&\frac{1}{2}({V}^{L,R}_{1}+\sqrt{2}{V}^{L,R}_{2}+{V}^{L,R}_{3}),
\nonumber \\
\bar{V}^{L,R}_{2}&=&\frac{1}{\sqrt{2}}({V}^{L,R}_{1}-{V}^{L,R}_{3}), \nonumber \\
\bar{V}^{L,R}_{3}&=&\frac{1}{2}({V}^{L,R}_{1}-\sqrt{2}{V}^{L,R}_{2}+{V}^{L,R}_{3}).
\label{eq-15}
\end{eqnarray}
Eqs. (\ref{eq-5}) and (\ref{eq-15}) give us interesting insight into
the transmission properties of the molecule. It is straightforward
to see that for some specific values of the magnetic flux and the
dot-lead matrix elements, the coupling between one or more molecular
states with the leads may vanish, giving rise to the formation of a
BIC. In particular, if $V_1=V_3$,
\begin{eqnarray}
\bar{V}^{L,R}_{1}&=&2V_{1}\cos\frac{\phi}{4}+\sqrt{2}V_{2},
\nonumber \\
\bar{V}^{L,R}_{2}&=&\mp i \sqrt{2}V_{1}\sin{\frac{\phi}{4}}, \nonumber \\
\bar{V}^{L,R}_{3}&=& 2V_{1}\cos\frac{\phi}{4}-\sqrt{2}V_{2}.
\label{eq-15B}
\end{eqnarray}
So that when $\phi=4n\pi$ ($n$ integer), the matrix elements
between the molecular state 2 and the left and right leads,
$\bar{V}^{L,R}_{2}$, cancel and such a state becomes a BIC. If it
also occurs that $V_1=V_2$, and $n$ is an odd multiple of $\pi$,
$\bar{V}^{L,R}_{2}\neq0$ but either $\bar{V}^{L,R}_{1}$ or
$\bar{V}^{L,R}_{3}$ vanish, occurring again a bound state in the
continuum. On the other hand, we can see of Eq. (\ref{eq-15B})
that if $V_1=V_3$ and $V_2=\sqrt{2}V_1$, two BICs are
simultaneously formed when $\phi=4n\pi$: one in the state 2
($\omega=\omega_0$), and other either in state 1
($\omega=\omega_0-\sqrt{2}t$) or 3 ($\omega=\omega_0+\sqrt{2}t$),
depending on the parity of $n$. In a parallel double quantum dot
molecule, a    simpler condition gives rise to one BIC, which is
formed whenever $\phi$ is a even multiple of $\pi$ (that is,
$\Phi=n\Phi_0$, $n$ integer).\cite{lldg,orellana} Notice that
bound states in the continuum occur for numerous combinations of
dot-lead couplings and Aharonov-Bohm phases, but, for simplicity,
we will restrict our attention to the particular cases: \emph{A.}
$V_1=V_2=V_3\equiv V$ and \emph{B.} $V_1=V_3\equiv V$ and
$V_2=\sqrt{2}V$. The equation of motion method in the molecular
basis $\{|\bar{1}\rangle,|\bar{2}\rangle,|\bar{3}\rangle\}$ gives
for the retarded Green's function in both cases
\begin{equation}
\bar{{\mathbf G}}^{r}(\omega)=\frac{1}{\Lambda}\left(
\begin{array}{lll}
\sqrt{2}t + \omega + i \bar{\Gamma}_{33}/2 &  \;\;\;\; 0 &  \;\;\;\; -i \bar{\Gamma}_{13}/2 \\
\;\;\;\; 0  & \Lambda/(\omega+i\bar{\Gamma}_{22}/2) & \;\;\;\; 0\\
-i \bar{\Gamma}_{31}/2 &  \;\;\;\; 0 &  -\sqrt{2}t + \omega + i \bar{\Gamma}_{11}/2 \\
\end{array}
\right), \label{eq-16}
\end{equation}
where
\begin{equation}
\Lambda=\frac{\bar{\Gamma}_{13}\bar{\Gamma}_{31}}{4}+(-\sqrt{2}t+\omega+i\frac{\bar{\Gamma}_{11}}{2})
(\sqrt{2}t+\omega+i\frac{\bar{\Gamma}_{33}}{2}), \label{eq-17}
\end{equation}
with $\bar{\Gamma}_{ij}=\bar{\Gamma}^L_{ij}+\bar{\Gamma}^R_{ij}$
($i,j=1,3$), where the matrices $\bar{\mathbf{\Gamma}}^L$ and
$\bar{\mathbf{\Gamma}}^R$ describe the tunneling between the
molecular states and the left and right leads, respectively.

\section{Conductance and density of states} \label{results}

\subsection*{A. $V_1=V_2=V_3$}

\begin{figure}
\begin{center}
  \includegraphics[width=7cm,angle=-90]{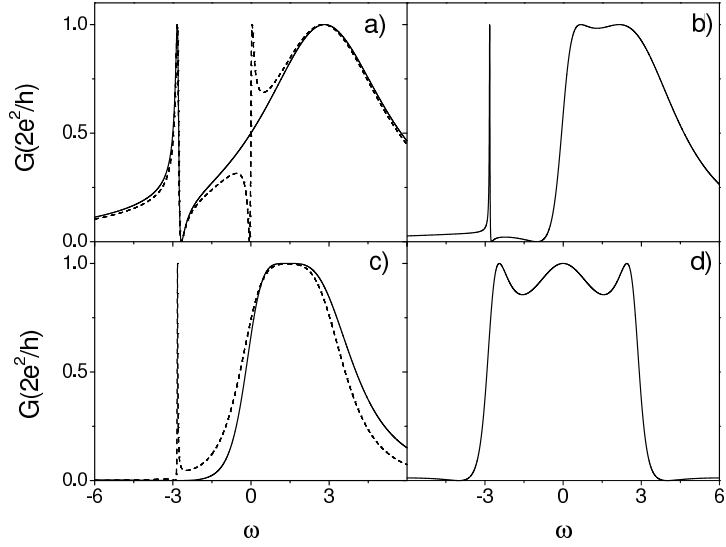}
\end{center}
\caption{Dimensionless conductance versus Fermi energy, in units of
$\gamma$, for $V_1=V_2=V_3=V$, $t=2$,  a) $\phi=0$ (solid line) and
$\phi=\pi/5$ (dash line), b) $\phi=3\pi/4$, c) $\phi=\pi$ (solid
line) and $\phi=6\pi/5$ (dash line) d) $\phi=2\pi$, for
$\omega_{0}=0$.} \label{fig2}
\end{figure}

When $V_1=V_2=V_3\equiv V$, the conductance takes the form
\begin{equation}
G(\omega)=\frac{2e^2}{h}\frac{\gamma^2[(\omega+2t\cos{\frac{\phi}{4}})^2-2(2t^2-\omega^2)
\cos{\frac{\phi}{2}}]^2} {[\omega
^2+\gamma^2(1-\cos{\frac{\phi}{2}})^2][(2t^2-\omega^2)^2+\gamma^2
(4t\cos{\frac{\phi}{4}}+ \omega[2+\cos{\frac{\phi}{2}}])^2]},
\label{ga}
\end{equation}
where $\gamma\equiv\Gamma^L_{ij}=\Gamma^R_{ij}$, for all
$i,j=1,2,3$. Figure \ref{fig2} shows the conductance as a function
of the Fermi energy for different values of $\phi$. In general,
three resonances are observed around the energies of the molecular
states. In Fig. \ref{fig2}(a), where $\phi=0$, the cancellation of
the resonance around $\omega=0$  accounts for the existence of the
BIC produced when $\phi=4\pi n$ ($n$ integer). The same figure
shows $G(\omega)$ for $\phi=\pi/5$, where the resonance
corresponding to $\omega=0$ is well resolved. For this value of
magnetic flux, as well as for arbitrary values of $\phi$, the
conductance displays two Fano antiresonances, at energies
\[
\omega=\frac{-\sqrt{2}t(\sqrt{2}\cos{\frac{\phi}{4}}\pm\sqrt{1+\cos\phi})}{1+2\cos{\frac{\phi}{2}}},
\]
as follows from Eq. (\ref{ga}). There are some special cases in
which the conductance shows only one antiresonance, namely, when
$\cos\phi=-1$ (i.e., when $\phi=(2n+1)\pi$, with $n$ integer), and
also when $\cos(\phi/2)=-1/2$ (that is, $\phi=2\pi(2n-2/3)$, $n$
integer), where one of the antiresonances goes to infinity and the
other keeps in a finite energy. We notice in Figs. \ref{fig2}(b) and
(c) (dashed line) that the resonance associated to the bonding state
remains very narrow for a wide range of values of $\phi$,
approximately from $\phi=-3\pi/2$ to $\phi=3\pi/2$, with a
periodicity $8\pi$. In the particular case when $\phi=2\pi (4n+1/2)$
($n$ integer), for instance in the solid line in (c), the resonance
is absent, so that the bonding state is a BIC. This is consistent
with Eqs. (\ref{eq-15B}), where $\bar{V}^{L}_3$ and $\bar{V}^{R}_3$
vanish for these values of $\phi$, leaving such a state decoupled of
the leads. In the same figure we included the curve for
$\phi=1.2\pi$ (dash line), to show the presence of that resonance
again. In (d), where $\phi=2\pi$ ($\Phi=\Phi_0$), the conductance is
totally symmetric with respect to $\omega=0$. These features repeat
whenever $\phi$ is an odd integer of $2\pi$. When $\phi$ is greater
than $2\pi$ the conductance spectrum suffers a reflection respect to
$\omega=0$, and every $4\pi$ (or two flux quanta) the roles of the
bonding and antibonding states are interchanged. Bound states in the
continuum are formed in the antibonding state when $\phi=2\pi
(4n+3/2)$ ($n$ integer).

\begin{figure}[h]
\begin{center}
  \includegraphics[width=7cm,angle=-90]{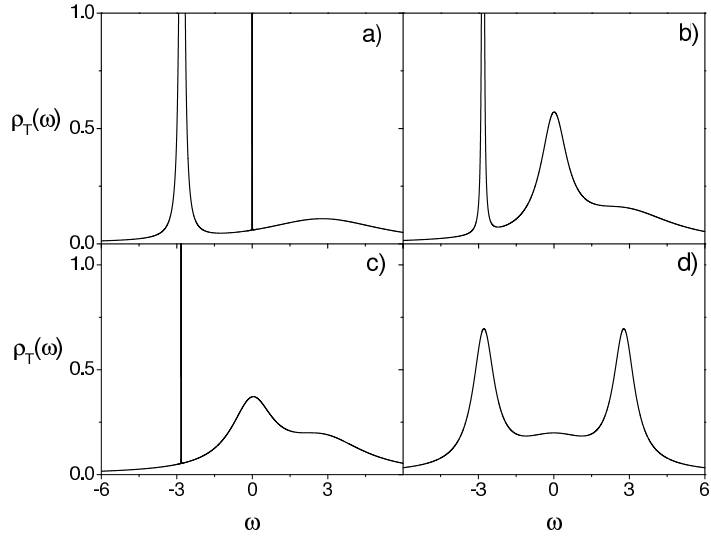}
\end{center}
\caption{Density of state versus Fermi energy, in units of
$\gamma$, for $V_1=V_2=V_3=V$, $t=2$, a) $\phi=0$, b)
$\phi=3\pi/4$, c) $\phi=\pi$ and d) $\phi=2\pi$, and $\omega_0=0$}
\label{fig3}
\end{figure}

Let us now examine the total density of states in the different
regimes. This has the form
\begin{equation}
\rho_T=\frac{\gamma}{\pi}\left[\frac{8t\omega\cos{\frac{\phi}{4}}+
(2t^2+\omega^2)(2+\cos{\frac{\phi}{2}})}{(2t^2-\omega^2)^2+\gamma^2
(4t\cos{\frac{\phi}{4}}+
\omega(2+\cos{\frac{\phi}{2}}))^2}+\frac{2\sin^2{\frac{\phi}{4}}}
{\omega^2+4\gamma^2\sin^4{\frac{\phi}{4}}}\right].
\end{equation}
Figure \ref{fig3} shows $\rho_T(\omega)$ for the same parameters of
Fig. \ref{fig2}. In the curves (a) and (c), where $\phi=0$ and
$\phi=\pi$, respectively, the delta functions corresponding to the
BICs are observed, superimposed to the peaks associated to the other
two molecular states with finite widths.

In the approximate range $2(4n-3/4)\pi\leq\phi\leq 2(4n+3/4)\pi$
($n$ integer) the total density of states can be approximated by the
sum of three Lorentzians at the energies
$\omega_-=-t(8\sqrt{2}t^2+4\gamma^2\cos(\phi/4)[2+\cos{(\phi/2)}])/(8t^2+\gamma^2[2+\cos{(\phi/2)}])$,
$0$ and $\sqrt{2}t$,
\begin{equation}
\rho_T\approx\frac{C}{\pi}\left[
\frac{\Gamma_-}{\Gamma_-^2+(\omega-\omega_-)^2} +
\frac{\Gamma_+}{\Gamma_+^2+(\omega-\sqrt{2}t)^2}\right]+\frac{1}{\pi}\frac{\Gamma_0}
{\Gamma_0^2+\omega^2}, \label{aprox1}
\end{equation}
where
\begin{eqnarray}
\Gamma_-&=&2\sqrt{2}t^2\gamma\frac{-4\cos{\frac{\phi}{4}}+\sqrt{2}[2+\cos{\frac{\phi}{2}}]}{8t^2
+\gamma^2[2+\cos{\frac{\phi}{2}}]^2},\\
\Gamma_0&=&2\gamma\sin^2{\frac{\phi}{4}}, \\
\Gamma_+&=&\gamma\frac{4\cos{\frac{\phi}{4}}+\sqrt{2}
[2+\cos{\frac{\phi}{2}}]}{2\sqrt{2}}, \label{w1}
\end{eqnarray}
and
\begin{equation}
C=\frac{1}{2}+\frac{\sqrt{2}\cos\frac{\phi}{4}}{2+\cos\frac{\phi}{2}}.
\end{equation}

\begin{figure}
\begin{center}
  \includegraphics[width=7cm,angle=-90]{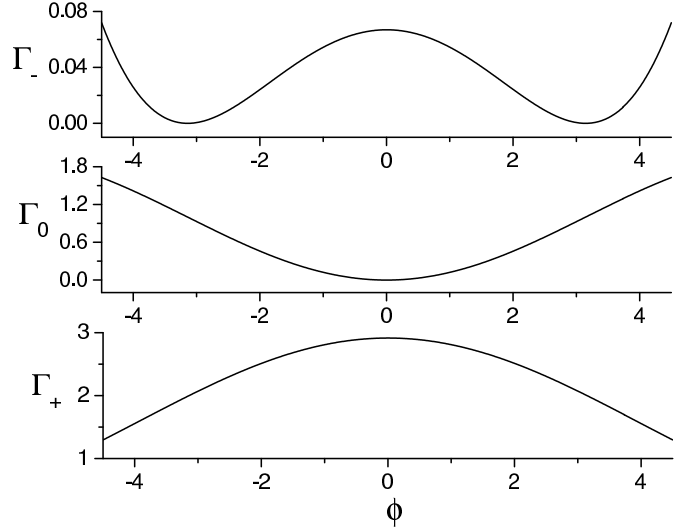}
\end{center}
\caption{Broadenings of the molecular states $\Gamma_-$, $\Gamma_0$
and $\Gamma_+$ as a function of $\phi$, for $V_1=V_2=V_3=V$ and
$t=2$.}\label{fig4}
\end{figure}

Figure \ref{fig4} shows the broadenings of the molecular states,
$\Gamma_-$, $\Gamma_0$ and $\Gamma_+$, (in units of $\gamma$) as a
function of $\phi$, for $t=2$, within the range of validity of the
approximations (\ref{w1}). The top curve gives account for the
formation of a BIC when $\phi$ is an odd integer of $\pi$, and shows
that the corresponding molecular state keeps very slightly coupled
to the leads for a wide range of the Aharanov-Bohm phase. In all the
interval, $\Gamma_-$ keeps smaller than a $7$ per cent of the level
broadening of a single dot, and in the interval $(\pi-1/2,\pi+1/2)$
it does not exceed a $1$ per cent. The middle figure shows that the
molecular state of intermediate energy ($\omega=0$) is a BIC when
$\phi=0$. The width of this molecular state keeps smaller than $1$
for all $\phi \in (-\pi,\pi)$ and is more sensitive to variations of
$\phi$ than the one formed in the bonding energy (top figure).
However, it is worth to notice that this long-lived state presents
for all phases in $(-4\pi/3,4\pi/3)$ larger lifetimes than the
long-lived state arising in the parallel-coupled double quantum dot
molecule symmetrically connected to leads (where $\Gamma=2\gamma
\sin^2{\phi/2}$).\cite{orellana} The behavior of the broadening of
the short-lived state is displayed in the bottom plot. This reaches
its maximum value (shortest lifetime) when $\phi=0$, where
$\Gamma_+\approx 2.91$.

\subsection*{B. $V_1=V_3$, $V_2=\sqrt{2}V$}

\begin{figure}
\begin{center}
  \includegraphics[width=7cm,angle=-90]{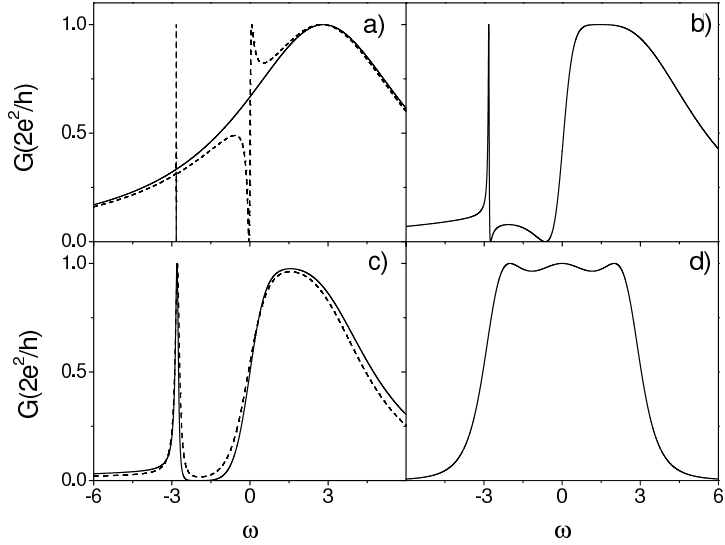}
\end{center}
\caption{Dimensionless conductance versus Fermi energy, in units of
$\gamma$, for $V_1=V_3=V$, $V_2=\sqrt{2}V$, $t=2$, a) $\phi=0$
(solid line) and $\phi=\pi/5$ (dash line),  b) $\phi=3\pi/4$, c)
$\phi=\pi$ (solid line) and $\phi=6\pi/5$ (dash line), d)
$\phi=2\pi$, and $\omega_{0}=0$.} \label{fig5}
\end{figure}

Interesting features in the electronic transmission take place when
the dot-lead couplings have the form $V_1=V_3=V$ and
$V_2=\sqrt{2}V$. Here the linear conductance reduces to
\begin{equation}
G=\frac{2e^2}{h}\frac{4\gamma^2[(t+\sqrt{2}\omega\cos{\frac{\phi}{4}})^2-t^2\cos{\frac{\phi}{2}}]^2}
{[\omega^2+\gamma^2(1-\cos{\frac{\phi}{2}})^2]
[(2t^2-\omega^2)^2+\gamma^2(4\sqrt{2}t\cos{\frac{\phi}{4}}+
\omega(3+\cos{\frac{\phi}{2}}))^2]}, \label{gb}
\end{equation}
where $\gamma\equiv=\Gamma^L_{ij}=\Gamma^R_{ij}$, for $i,j=1,3$.
Fig. \ref{fig5} shows the conductance for different values of the
Aharonov-Bohm phase. For $\phi=0$ (Fig. \ref{fig5}(a)) this
exhibits a single resonance around the antibonding energy, so that
the two other molecular states are bound states in the continuum.
This also follows of Ecs. (\ref{eq-15B}), since $\bar{V}^{L,R}_2$
as well as $\bar{V}^{L,R}_3$ cancel whenever $\phi=8n\pi$ ($n$
integer). Also, it can be seen that the roles of the bonding and
antibonding states are interchanged every $4\pi$, and that the
antibonding and the molecular state of intermediate energy both
collapse to BICs when $\phi=4\pi(2n-1)$ ($n$ integer).
For arbitrary values of the Aharonov-Bohm phase the spectrum
presents three resonances, and a number of Fano antiresonances that
oscillates between two and zero. From Eq. (\ref{gb}) we note that
the conductance is zero at
\begin{equation}
\omega=\frac{-\sqrt{2}t[1\pm\sqrt{\cos(\phi/2)}]}{2\cos{\phi/4}},
\end{equation}
where we see that when $(4n-1)\pi<\phi<(4n+1)\pi$ ($n$ integer)
there are two antiresonaces, as shown by Fig. \ref{fig5}(b) where
$\phi=3\pi/4$. When $\phi=n\pi$ ($n$ odd) only one point of zero
conductance exists, as observed in (c). For
$(4n-3)\pi<\phi<(4n-1)\pi$ ($n$ integer) the numerator of Eq.
(\ref{gb}) is complex and the conductance does not exhibit
antiresonances (Figs. (c), dashed line, and (d)). In (d) $G(\omega)$
is symmetrical around $\omega=0$, and has the form
\[
G=\frac{2e^2}{h}\frac{16
\gamma^2t^4}{(4\gamma^2+\omega^2)(4t^4-4(t+\gamma)(t-\gamma)\omega^2+\omega^4)},
\]
which corresponds exactly to the conductance of a triple quantum
dot molecule connected in series.\cite{zeng} It is interesting to
note that if the dots are not coupled directly, that is, $t=0$,
the transmission is suppressed for all energies (perfect
reflector). An analogous result is found in two parallel quantum
dots in presence of a magnetic field.\cite{kubala} This situation
never occurs in the triple molecule when the dot-lead couplings
are equal.

The total density of states is given by
\begin{equation}
\rho_T(\omega)=\frac{\gamma}{\pi}\left[
\frac{8\sqrt{2}t\omega\cos{\frac{\phi}{4}}+(2t^2+\omega^2)(3+\cos{\frac{\phi}{2}})}
{(2t^2-\omega^2)^2+\gamma^2[4\sqrt{2}t\cos{\frac{\phi}{4}}+
\omega(3+\cos{\frac{\phi}{2}})]^2}
+\frac{2\sin^2{\frac{\phi}{4}}}{\omega^2+4\gamma^2\sin^4{\frac{\phi}{4}}}
\right]. \label{dos2}
\end{equation}

In the range $2(4n-3/8)\pi\leq\phi\leq 2(4n+3/8)\pi$ ($n$ integer),
$\rho_T$ can be approximated by a sum of Lorentzians of the form Eq.
(\ref{aprox1}), where
$\omega_-=-4\sqrt{2}t(2t^2+\gamma^2\cos(\phi/4)[3+\cos{(\phi/2)}])/
(8t^2+\gamma^2[3+\cos(\phi/2)]^2)$,
\begin{equation}
C=\frac{4\cos^4{\frac{\phi}{4}}}{3+\cos\frac{\phi}{2}}
\end{equation}
and the broadenings are given by
\begin{eqnarray}
\Gamma_-&=&8\gamma
\frac{t^2\sin^4{\frac{\phi}{8}}}{2t^2+\gamma^2(1+\cos^2{\frac{\phi}{4}})^2} \\
\Gamma_0&=& 2\gamma\sin^2\frac{\phi}{4}\\
\Gamma_+&=&4\gamma\cos^4{\frac{\phi}{8}} \label{width2}
\end{eqnarray}

\begin{figure}
\begin{center}
  \includegraphics[width=7cm,angle=-90]{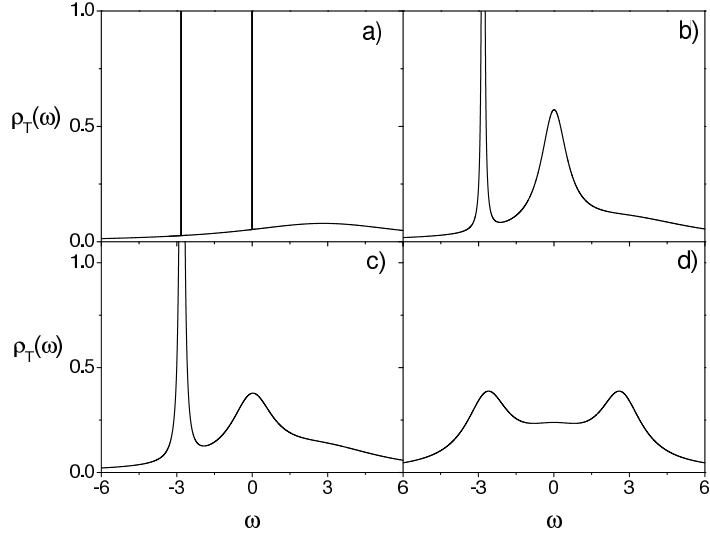}
\end{center}
\caption{Density of state versus Fermi energy, in units of $\gamma$,
for  $V_1=V_3=V$, $V_2=\sqrt{2}V$, $t=2$, a) $\phi=0$, b)
$\phi=3\pi/4$, c) $\phi=\pi$ and d) $\phi=2\pi$, for
$\omega_{0}=0$.} \label{fig6}
\end{figure}
Fig. \ref{fig6} shows $\rho_T$ for the same parameters of
Fig.\ref{fig5}. In figure Fig.\ref{fig6}(a), where $\phi=0$, the
density of states is the superposition of two Dirac delta's
localized at $\omega=\omega_{-}$ and $\omega=0$ (corresponding to
the BICs) plus a Lorentzian at $\omega=\omega_{+}$ with width
$4\gamma$. When $\phi=2\pi$, as in (d), the density of states
correspond to that of a triple molecule connected in series.
\begin{figure}
\begin{center}
  \includegraphics[width=7cm,angle=-90]{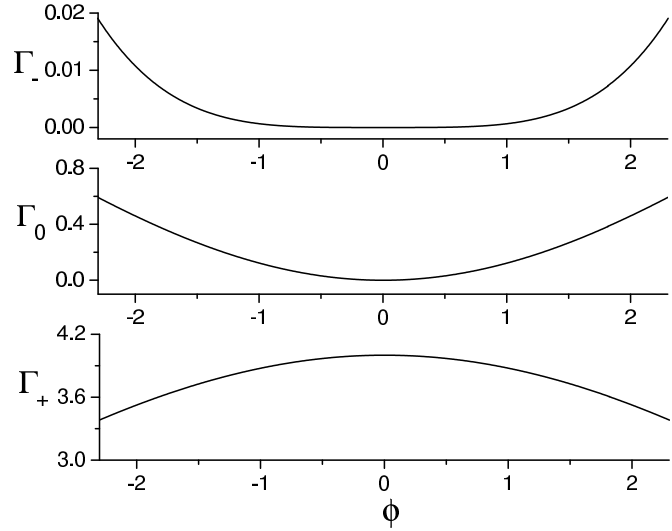}
\end{center}
\caption{Broadenings of the molecular states $\Gamma_-$, $\Gamma_0$
and $\Gamma_+$ as a function of $\phi$, for $V_1=V_3=V$,
$V_2=\sqrt{2} V$ and $t=2$.} \label{fig7}
\end{figure}

Fig. \ref{fig7} displays the broadenings $\Gamma_-$, $\Gamma_0$
and $\Gamma_+$ as a function of $\phi$, for $t=2$, in the range of
validity of Eqs. (\ref{width2}). In the top plot shows the BIC
formed in the bonding state when $\phi=0$ and the robustness of
the such a long-lived state against variations of the magnetic
field. Notice that $\Gamma_-$ remains smaller than $0.02$ in all
the plotted range of $\phi$, and that is very close to zero in a
wide interval around $\phi=0$. For instance, for $\phi \in
(-\pi/3,\pi/3)$, $\Gamma_-$ keeps smaller than $8\times 10^{-4}$,
that is, less than $0.08$ per cent of the level broadening of a
single quantum dot. The broadening $\Gamma_0$ (which coincide with
$\Gamma_0$ in the previous setting) is more sensitive to
variations of the magnetic field than $\Gamma_-$, as shown in the
middle figure.

The robustness of the molecular states for the triple quantum dot
molecule can be understood physically as follows: in the triple
molecule the phases acquired by the electron in covering some of
the paths, namely, those containing the central quantum dot, are
smaller than those gained when the electron travel though paths
containing the outer dots. For instance, along the central path
the electron does not accumulate any phase. This information is
contained in the effective couplings of the molecular states with
the leads, and therefore their dependence on the Aharonov-Bohm
phase is less sensitive in comparison to the double quantum dot
case.

We interpret the formation of BICs in this system in the same
sense of reference [\onlinecite{friedrich}], that is, as result of
the quantum interference between resonances of different channels
through the multiple connected quantum dots. The levels of the
quantum dots are hybridized through the common leads, forming
these states of infinite lifetimes. In a more realistic model on
transmission through the quantum dot molecule one should take into
account the electron-electron interaction. In recent works some
authors have found results that can be interpreted as BICs that
survive to the interaction effects. For instance, Ding {\em et
al.}\cite{ding} study a parallel double quantum dot in the Kondo
regime by using the finite-U slave boson technique, and found a
$\delta$-peak structure in the density of state for the energies
inside the band, signal of a bound state in the continuum. On the
another hand, Busser {\em et al.}\cite{busser} study the transport
properties of multilevel quantum dots in Kondo regime and report
the formation of localized states. We interpret these results as
BICs in the same sense described in the present paper. However, we
think that the above results are not conclusive and further
research is necessary to know the effect of the electron-electron
interaction on the formation of BICs.

\section{Larger molecules}
\label{Ndots}

It is natural to ask about the existence of bound states in the
continuum in molecules of $N$ quantum dots, with $N$ arbitrary. As
seen for the double\cite{lldg} and triple parallel-coupled
molecules, the existence a magnetic field is not essential for the
formation of BICs , but it is needed just a certain relation of
symmetry between the couplings between dots and leads for these to
take place. In fact, the maximum number of simultaneous bound
states may occur when the magnetic flux is zero. So, a first
approach to the problem of a parallel-coupled molecule of $N$
quantum dots can be obtained by assuming that there is no field
present.

The $j$-th component of the eigenfunction $|\psi_n\rangle$ of a
linear chain consisting of $N$ identical quantum dots with
energies $\varepsilon_0$ and tunnel coupling between dots $t$ is
given by
\begin{equation}
\psi_{j,n}=\sqrt{\frac{2}{N+1}}\sin{\frac{jn\pi}{N+1}},\quad
j=1,2,\dots,N \label{components}
\end{equation}
and the corresponding eigenenergy is
\begin{equation}
E_n=\varepsilon_0-2t\cos{\frac{n\pi}{N+1}}.
\end{equation}
The Hamiltonian describing the molecule-leads interaction is
\begin{equation}
H_{i}=\sum_{i=1}^{N}\sum_{k\in L}(V_{i}d^{\dag}_ic_k+\mbox{
h.c.})+\sum_{i=1}^{N}\sum_{k\in R}(V_{i}d^{\dag}_ic_k+\mbox{
h.c.})
\end{equation}
where we have assumed that $V^L_i=V^R_i\equiv V_i$, for all
$i=1,2,\dots,N$. To search for conditions for the formation of BICs,
we look at the couplings of the molecular states with the leads,
$\bar{V}_n$ ($n=1,2,\dots,N$). These can be obtained by the
transformation
\begin{equation}
\mathbf{\bar{V}}=P \mathbf{V}, \label{vpv}
\end{equation}
where $\mathbf{\bar{V}}$ y $\mathbf{V}$ are column vectors with
elements $\bar{V}_n$ and $V_n$ ($n=1,2,\dots,N$), respectively,
and $P$ is the matrix composed of eigenvectors $|\psi_n\rangle$
($n=1,2,\dots,N$), which components are given by
(\ref{components}). Namely,
\begin{equation}
P=\sqrt{\frac{2}{N+1}}\left( \begin{array}{ccccc}
\sin{\frac{\pi}{N+1}} & \sin{\frac{2\pi}{N+1}} & \sin{\frac{3\pi}{N+1}} & \cdots & \sin{\frac{N\pi}{N+1}} \\
\sin{\frac{2\pi}{N+1}} & \sin{\frac{4\pi}{N+1}} & \sin{\frac{6\pi}{N+1}}&  \cdots & \sin{\frac{2N\pi}{N+1}}\\
\vdots& \vdots & \vdots & \ddots &  \vdots\\
\sin{\frac{N\pi}{N+1}} & \sin{\frac{2N\pi}{N+1}} &
\sin{\frac{3N\pi}{N+1}} & \cdots & \sin{\frac{N^2\pi}{N+1}}
\end{array}\right).
\end{equation}
It is direct to show that the matrix elements of $P$ have the
following properties:
\begin{eqnarray}
P_{i,j}&=&P_{i,N-j+1} \quad \mbox{$i$ odd}\label{prop1} \\
P_{i,j}&=&-P_{i,N-j+1} \quad \mbox{$i$ even} \label{prop2} \\
P_{i,(N+1)/2}&=&0, \quad \mbox{$N$ odd and $i$ even} \label{prop3}
\end{eqnarray}
These relations, together with Eq. (\ref{vpv}), allow us to get
information on some symmetries leading to the formation of BICs.
For instance, if there is up-down symmetry, that is,
\[
V_{i}=V_{N-i+1}, \quad \mbox{$i=1,2,\cdots,N$},
\]
then by the condition (\ref{prop2}),
\[
\bar{V}_j=0, \quad \mbox{for all $j$ even}.
\]
Thus, if $N$ is even, $N/2$ bound states in the continuum are
ensured. In turn, if $N$ is odd, both conditions (\ref{prop2}) and
(\ref{prop3}) are simultaneously required for the formation of
$(N-1)/2$ bound states in the continuum.

We illustrate the above analysis by considering in detail the
cases $N=4$ and $N=5$. We get conditions for the formation of
additional BICs in each example. For the molecule of four quantum
dots, Eqn. (\ref{vpv}) reduces to
\begin{eqnarray}
\bar{V}_{1}&=&\alpha_+(V_2+V_3)+\alpha_-(V_1+V_4) \nonumber \\
\bar{V}_{2}&=&\alpha_-(V_2-V_3)+\alpha_+(V_1-V_4) \nonumber \\
\bar{V}_{3}&=&-\alpha_-(V_2+V_3)+\alpha_+(V_1+V_4)\nonumber \\
\bar{V}_{4}&=&-\alpha_+(V_2-V_3)+\alpha_-(V_1-V_4),
\end{eqnarray}
where $\alpha_\pm=(1/2)\sqrt{1\pm1/\sqrt{5}}$. We see that if
$V_1=V_4$ and  $V_2=V_3$  both $\bar{V}_2$ and $\bar{V}_4$ are
canceled, thus occurring two BICs. We notice that if also
$\alpha_+V_2=\mp\alpha_-V_1$, either $\bar{V}_{1}$ or
$\bar{V}_{3}$ vanishes, having three of the four molecular states
decoupled from the continuum.

For a molecule of five quantum dots we have
\begin{eqnarray}
\bar{V}_1&=&\frac{1}{2}[V_2+V_4+\frac{\sqrt{3}}{3}(V_1+2V_3+V_5)]\nonumber \\
\bar{V}_2&=&\frac{1}{2}(V_1+V_2-V_4-V_5) \nonumber \\
\bar{V}_3&=&\frac{\sqrt{3}}{3}(V_1-V_3+V_5)\nonumber \\
\bar{V}_4&=& \frac{1}{2}(V_1-V_2+V_4-V_5) \nonumber \\
\bar{V}_5&=&\frac{1}{2}[-(V_2+V_4)+\frac{\sqrt{3}}{3}(V_1+2V_3+V_5)].
\end{eqnarray}
Because of condition (\ref{prop3}) neither $\bar{V}_2$ nor
$\bar{V}_4$ depends on $V_3$. This, together with $V_1=V_5$ and
$V_2=V_4$ lead to the occurrence BICs in the molecular states 2
and 3. Up to two new bound states may arise if the following
conditions are met: $V_3=2V_1$ suppresses $\bar{V}_3$ and this
together with $V_2=\mp V_1$ cancel either $\bar{V}_1$ or
$\bar{V}_5$.

\section{Conclusions}\label{conclusions}

We have investigated electron transport through a parallel-coupled
triple quantum dot molecule in presence of a magnetic field. The
conductance spectrum exhibits a Breit-Wigner and two Fano
resonances, which positions and widths are controlled by the
magnetic field. Every two flux quanta ($\phi=4\pi$), the roles of
the bonding and antibonding states are interchanged. We have
examined the dependence and broadenings of the molecular states as
a function of the magnetic flux for two different sets of
parameters, finding that several regimes of transmission are
possible, including the formation of extremely narrow resonances
and bound states in the continuum. We have shown that by
manipulating the symmetries of the system, up to two simultaneous
BICs can be formed. We restricted our analysis to systems with
up-down and left-right symmetries. In Ref. \onlinecite{lldg} it is
shown that the breaking of the first of those symmetries hinders
the formation of BICs, but states of very long lifetimes still
occur. With respect to the left-right symmetry, this is not
essential in the existence of BICs, as demonstrated in Ref.
\onlinecite{orellana} for the double quantum dot molecule. We
extended the study to molecules of $N$ quantum dots in the absence
of magnetic field, finding that the up-down symmetry ensures the
occurrence of $N/2$ BICs for $N$ even, and of $(N-1)/2$ for $N$
odd. Additional conditions are required for the existence of a
larger number of simultaneous bound states in the continuum, as
shown for molecules of four and five dots. In both cases up to
$N-1$ BICs may exist at the same time, $N$ being the number of
dots.

The possibility of having molecular states decoupled from the
leads, and the fact that these states are controllable by an
external magnetic field and gate potentials, are interesting
features of the studied system which might be useful to classical
information theory. Two orthogonal stable states of the molecule,
that is, two simultaneous BICs, could be used as microscopic units
for storing information (classical bits). Storage of quantum
information requires a complete stable plane in the Hilbert space
of the molecule.\cite{nielsen} Quantum dot molecules seem to be
suitable systems to study BICs experimentally, because of the
possibility of controlling parameters. In fact, other quantum
interference phenomena have been demonstrated in this kind of
systems, such as Fano\cite{kobayashi} and
Aharonov-Bohm\cite{yacobi} effects.

\section*{Acknowledgments}
We thank Luis Roa for useful comments. This work acknowledges
financial support from FONDECYT, under grants 1040385 and 1020269.
M. L. L. de G. receives financial support from Milenio ICM
P02-049-F and P. A. O. from Milenio ICM P02-054-F.

\end{document}